\documentclass[submission,copyright,creativecommons]{eptcs}
\providecommand{\event}{FOCLASA 2010} % Name of the event you are submitting to
\usepackage{breakurl}             % Not needed if you use pdflatex only.

\usepackage{graphicx}
\usepackage[centertags]{amsmath}
\usepackage{amsfonts}
\usepackage{amssymb}
\usepackage{alltt}
\usepackage{moreverb}
\usepackage{paralist}
\usepackage{fancybox}
\usepackage{boxedminipage}
\usepackage{enumerate}
\usepackage{url}
\usepackage[all,2cell,ps]{xy} %%xypic
\usepackage{subfigure}

\usepackage[francais]{minitoc}

\newtheorem{proposition}{Proposition}[section]

\newtheorem{definition}[proposition]{Definition}

\newtheorem{theorem}[proposition]{Theorem}
\newtheorem{lemma}[proposition]{Lemma}

\newtheorem{example}[proposition]{Example}
\title{On Coordinating Collaborative Objects}
\author{Abdessamad Imine
\institute{Nancy University and INRIA Grand-Est\\ Nancy, France}
\email{abdessamad.imine@loria.fr}
}
\def\titlerunning{On Coordinating Collaborative Objects}
\def\authorrunning{A. Imine}
\begin{document}
\maketitle

\sloppy

\begin{abstract}
A collaborative object represents a data type (such as a text document) 
designed  to be  shared by  a group of dispersed users.
The Operational Transformation (OT)
is a coordination approach used for supporting optimistic replication for these
objects. It allows
the users to concurrently update the shared data and exchange their updates
in any order since the  convergence of all replicas, \textit{i.e.}
the fact  that all users view the same data, is ensured in all cases. 
However, designing
algorithms for achieving convergence with the OT approach is a critical and
challenging issue.
In  this paper,  we propose  a formal compositional
method  for specifying  complex  collaborative objects.
The most important  feature of our method is  that designing an OT algorithm
for the  composed  collaborative  object   can  be  done  by  reusing  the
OT algorithms of component collaborative objects.  By using our method,
we  can  start from  correct  small  collaborative  objects which  are
relatively easy to handle and incrementally combine them to build more
complex collaborative objects.

\noindent{\textbf{Key words:} Collaborative Editors, Operational Transformation,
Component-based design, Algebraic Specifications.}
\end{abstract}

\section{Introduction}

\noindent{\textbf{Motivation.}}
Collaborative editors constitute  a class of distributed systems where
dispersed users interact by manipulating simultaneously some
shared objects like texts, images, graphics, etc. 
%One of the main challenges is the data consistency. 
To improve data availability, 
%optimistic consistency control techniques are commonly used. 
the shared data is replicated so that the users update their local data
replicas and exchange their updates between them. So, the updates are applied in different orders 
at different replicas of the object. This potentially leads to divergent 
(or different) 
replicas -- an undesirable situation for collaborative editors.
\emph{Operational Transformation} (OT) is an  optimistic technique
which has been proposed to overcome the divergence problem~\cite{Ellis89}. This technique
consists of an algorithm which transforms an update (previously
executed by some other user) according to local concurrent ones in
order to achieve convergence. 
It is used in many
collaborative editors including 
%Joint Emacs~\cite{Ressel.ea:96} (an
%Emacs collaborative editor), 
CoWord~\cite{Sun06} and CoPowerPoint~\cite{Sun06} (a collaborative
version of MicroSoft Word and PowerPoint respectively),  and the Google Wave 
(a new Google 
platform\footnote{\url{http://www.waveprotocol.org/whitepapers/operational-transform}}).

It should be noted that the data consistency relies crucially on the 
correctness of an OT
algorithm. According to ~\cite{Ressel.ea:96},  the consistency is ensured iff
the OT algorithm satisfies two convergence
properties $TP1$ and $TP2$ (that will be detailed in Section $2$). 
Finding such an algorithm and 
proving that it satisfies $TP1$ and $TP2$ is not an easy task because it requires
analyzing a large number of situations. Moreover,
when we  consider a
complex object  (such as a filesystem  or an XML  document that are composite
of several primitive objects) the formal
design of its  OT algorithm becomes very tedious  because of the large
number of  updates and coordination situations to  be considered if
we  start from  scratch.

\medskip
\noindent{\textbf{Related Work.}}
Research efforts have been focused on automatically verifying the correctness of
OT algorithms by using either a theorem prover~\cite{Imi06} or a 
model-checker tool~\cite{Han09}. To the best of our knowledge,~\cite{Imi07} is the
first work that addresses the formal compositional design of OT algorithms.
In this work, two \emph{static} constructions (where the number of objects 
to combine is fixed) 
have been proposed for composing collaborative objects:
\begin{inparaenum}[(i)]
\item the first construction has as a basic semantic property to combine
      components without allowing these components to interact;
\item as for the second one it enables components to communicate by means of a
      shared part.
\end{inparaenum}

\medskip
\noindent{\textbf{Contributions.}}
As continuation of ~\cite{Imi07},
we propose  in this paper how to combine an
arbitrary number of collaborative objects by
using a dynamic composition in such a way the objects are created and
deleted dynamically.
The most important  feature of our method is  that designing an OT algorithm
for the  composed  collaborative  object   can  be  done  by  reusing  the
OT algorithms of component collaborative objects.
%The most important  feature of our method is  that the verification of
%the  composed  collaborative  object   can  be  done  by  reusing  the
%verification of component collaborative objects.  
By using our method,
we  can  start from  correct  small  collaborative  objects 
(\textit{i.e.}, they satisfy convergence properties) which  are
relatively easy to handle and incrementally combine them to build more
complex collaborative objects that are also correct.

\medskip
\noindent{\textbf{Roadmap.}}
This paper is organized as follows:
in Section~\ref{sec:ot} we give the basic concepts of the OT approach.
The ingredients of our formalization for specifying the collaborative object and OT
algorithm are given in Section~\ref{sec:co}.
In Section~\ref{sec:dc}, we present how to specify the dynamic composition of
collaborative objects in algebraic framework. Section~\ref{sec:cor} gives the
correctness of our dynamic composition approach.
Finally, we give conclusions and present future work.

\section{Operational Transformation Approach}\label{sec:ot}
Due to high  communication latencies in wide-area  and  mobile  
wireless  networks the replication of collaborative objects is commonly 
used in distributed collaborative systems. But this choice is not
without problem as we will see in next sub-section.%\vspace{-2mm}

\subsection{Convergence Problems}
One  of the  significant  issues when  building collaborative
editors with a replicated  architecture and an arbitrary communication
of messages between users  is the \textit{consistency maintenance} (or
\textit{convergence})  of all  replicas. To  illustrate  this problem,
consider the following example:

\begin{example}\label{exmp:e11}
  Consider   the   following group   text   editor   scenario   (see
  Figure~\ref{fig:incons}): there  are two users (sites)  working on a
  shared  document  represented by  a  sequence  of characters.  These
  characters  are addressed  from  $0$ to  the  end of  the document.  
  Initially,  both copies hold  the string  ``\texttt{efecte}''.  User
  $1$  executes  operation  $op_1 =  Ins(1,\mbox{``\texttt{f}''})$  to
  insert the character  ``\texttt{f}'' at position $1$.  Concurrently,
  user  $2$   performs  $op_2  =  Del(5)$  to   delete  the  character
  ``\texttt{e}''  at  position  $5$.   When  $op_1$  is  received  and
  executed   on   site   $2$,   it  produces   the   expected   string
  ``\texttt{effect}''.  But,  when $op_2$ is received on  site $1$, it
  does not take  into account that $op_1$ has  been executed before it
  and it produces the  string ``\texttt{effece}''.  The result at site
  $1$  is different  from the  result of  site $2$  and  it apparently
  violates  the   intention  of   $op_2$  since  the   last  character
  ``\texttt{e}'', which  was intended to be deleted,  is still present
  in the  final string.   
  %Consequently, we obtain  a \emph{divergence}
  %between sites $1$ and $2$.  It  should be pointed out that even if a
  %serialization protocol~\cite{Ellis89}  was used to  require that all
  %sites  execute $op_1$  and $op_2$  in the  same order  to  obtain an
  %identical result ``\texttt{effece}'', this identical result is still
  %inconsistent with the original intention of $op_2$.

\begin{figure}[t]%[htbp]
 \begin{minipage}[t]{0.5\linewidth}
\begin{small}
\centerline{\xymatrix@C=10pt@M=2pt@R=10pt{
*+[F-,]\txt{site 1 \\ ``efecte''} \ar@{.}'[d]'[dd]'[ddd][dddd] &
*+[F-,]\txt{site 2 \\ ``efecte''} \ar@{.}'[d]'[dd]'[ddd][dddd] \\
op_1=Ins(1,``f'') \ar[ddr]  & op_2=Del(5) \ar[ddl] |!{[l];[dd]}\hole \\
*+[F]{\txt{``effecte''}} & *+[F]{\txt{``efect''}} \\
  Del(5)     &   Ins(1,``f'') \\
*+[F]{\txt{``effece''}}  & *+[F]{\txt{``effect''}} \\
}}
\end{small}
  \caption{Incorrect integration.}
  \label{fig:incons}
 \end{minipage}
 \begin{minipage}[t]{0.5\linewidth}
\begin{small}
\centerline{\xymatrix@C=20pt@M=2pt@R=10pt{
*+[F-,]\txt{site 1 \\ ``efecte''} \ar@{.}'[d]'[dd]'[ddd][dddd] &
*+[F-,]\txt{site 2 \\ ``efecte''} \ar@{.}'[d]'[dd]'[ddd][dddd] \\
op_1=Ins(1,``f'') \ar[ddr]  & op_2=Del(5) \ar[ddl] |!{[l];[dd]}\hole \\
*+[F]{\txt{``effecte''}} & *+[F]{\txt{``efect''}} \\
 IT(op_2,op_1)=Del(6)     & Ins(1,``f'') \\
*+[F]{\txt{``effect''}}  & *+[F]{\txt{``effect''}} \\}}
\end{small}
  \caption{Integration with transformation.}
  \label{fig:transf}
 \end{minipage}

\end{figure}

\end{example}

To   maintain  convergence,   an   OT  approach   has  been   proposed
in~\cite{Ellis89}. It 
consists of application-dependent  transformation algorithm  such  that for  every
possible pair of concurrent updates, the application programmer has to
specify  how to  merge these  updates regardless  of  reception order.
We denote this
algorithm by a function $IT$, called 
\emph{inclusion transformation}~\cite{Sun.ea:98}.

\begin{example}
  In Figure~\ref{fig:transf},  we illustrate the effect of  $IT$ on the
  previous example. When $op_2$ is  received on site $1$, $op_2$ needs
  to    be    transformed   in order to include the effects of
  $op_1$:
  $IT((Del(5),Ins(1,\mbox{``\texttt{f}''}))  =  Del(6)$.  The  deletion
  position  of $op_2$  is incremented  because $op_1$  has  inserted a
  character at position $1$, which  is before the character deleted by
  $op_2$. Next,  $op'_2$ is  executed on site  $1$.  In the  same way,
  when $op_1$ is  received on site $2$, it  is transformed as follows:
  $IT(Ins(1,\mbox{``\texttt{f}''}),Del(5))                             =
  Ins(1,\mbox{``\texttt{f}''})$;  $op_1$   remains  the  same  because
  ``\texttt{f}'' is inserted before the deletion position of $op_2$.
%\end{example}
Intuitively we can write the transformation $IT$ as follows:
\begin{center}
\begin{footnotesize}
\begin{boxedverbatim}
IT(Ins(p1,c1),Ins(p2,c2)) = if (p1 < p2) return Ins(p1,c1)
                           else return Ins(p1+1,c1)
                           endif;
\end{boxedverbatim}
\end{footnotesize}
\end{center}
\end{example}
%%\vspace{-8mm}

\subsection{Transformation Properties}
Notation $[op_1; op_2; \ldots; op_n]$ represents an operation sequence.
We denote $Do(X, st) = st'$ when an operation (or an operation sequence) 
$X$ is executed
on a replica state $st$ and produces a replica state $st'$.

Using   an   OT   algorithm   requires   to   satisfy   two
properties~\cite{Ressel.ea:96}, called \emph{transformation properties}.   
Given  three  operations   $op$, $op_1$  and
$op_2$,   with  $op'_2=IT(op_2,op_1)$   and   $op'_1=IT(op_1,op_2)$,  the
conditions are as follows:
%%\vspace{-2mm}
\begin{itemize}
\item \textbf{Property $\mathbf{TP1}$}: $Do([op_1; op'_2], st)\,=\,
      Do([op_2; op'_1], st$), for every state $st$.
\item \textbf{Property $\mathbf{TP2}$}: 
        $IT(IT(op,op_1),op'_2)\,=\,IT(IT(op,op_2),op'_1)$.
\end{itemize}

$\mathbf{TP1}$ defines a  \emph{state identity} and ensures that  if $op_1$ and
$op_2$ are concurrent, the effect of executing $op_1$ before $op_2$ is
the  same  as  executing  $op_2$  before  $op_1$.  This  condition  is
necessary but not sufficient  when the number of concurrent operations
is greater than two.  As  for $\mathbf{TP2}$, it ensures that transforming $op$
along equivalent and different  operation sequences will give the same
result.   
%In  ~\cite{Suleiman.ea:98},  the  authors have  proved  that
%conditions 
Properties $\mathbf{TP1}$  and $\mathbf{TP2}$ are  sufficient to ensure  
the convergence
property for \textit{any number} of concurrent operations which can be
executed in \textit{arbitrary order}~\cite{Ressel.ea:96}.

\section{Primitive Collaborative Objects}\label{sec:co}
\subsection{Basic Notions}\label{bn}
In this sub-section we present terminology and notation that are used
in the following sections. We assume that the reader is familiar with
algebraic specifications. For more background on this topic see
\cite{Wir90,Gog00}.

A \emph{many-sorted  signature} $\Sigma$ is a
pair  $(S,F)$  where  $S$ is  a  set  of  \emph{sorts}  and $F$  is  a
$S^{*}\times  S$-sorted  set   (of  \emph{function  symbols}). Here,
$S^{*}$ is the  set of finite (including empty)  sequences of elements
of $S$.  Saying that $f: s_1\times\ldots\times s_n\rightarrow s$ is in
$\Sigma=(S,F)$ means  that $s_1\ldots\ s_n  \in S^{*}$, $s\in  S$, and
$f\in F_{s_1\ldots s_n ,s}$. A $\Sigma$-\emph{algebra} $A$ interprets
sorts as sets and operations as appropriately typed functions.
A \emph{signature morphism} $\Phi:\Sigma\rightarrow\Sigma'$ is a pair ($f$,$g$), 
such that $f:S\rightarrow S'$ and $g:\Sigma\rightarrow\Sigma'_{f^*,f}$
an ($S^{*}\times  S$)-sorted function. Usually, we ignore the distinction
between $f$ and $g$ and drop all subscripts, writing $\Phi(s)$ for
$f(s)$ and $\Phi(\sigma)$ for $g(\sigma)$ such that 
$\sigma\in F_{s_1\ldots s_n ,s}$. We denote the sort of booleans as \texttt{Bool}.

Let $X$ be a family of sorted variables and let $T_{\Sigma}(X)$ be
the algebra of $\Sigma$-terms. An
\emph{equation} is  a formula of the form  $l=r$ where $l$, 
$r\in T_{\Sigma}(X)_s$ for some sort $s\in S$. 
A \emph{conditional}
equation     is     a    formula     of     the    following     form:
$\bigwedge^{n}_{i=1}a_i=b_i   \implies  l=r$, where $a_i$, 
$b_i\in T_{\Sigma}(X)_{s_i}$. An  \emph{algebraic specification}  is  
a pair  $(\Sigma,E)$
where  $\Sigma$ is  a  many-sorted signature  and  $E$ is  a set of
(conditional) $\Sigma$-equations,    called   \emph{axioms}    of
$(\Sigma,E)$. A $(\Sigma,E)$-model is a $\Sigma$-algebra $A$ that satisfies
all the axioms in $E$. We write $A\models^{\Sigma} E$ to indicate that $A$ is
a $(\Sigma,E)$-model. 
Given a signature morphism $\Phi:\Sigma\rightarrow\Sigma'$ and a 
$\Sigma'$-algebra $A'$, the \emph{reduct} of $A'$ to $\Sigma$, denoted
$\Phi(A')$, represents carriers $A'_{\Phi(s)}$ for $s\in S$ and operations
$\sigma_{\Phi(s)}$ for $\sigma\in \Sigma_{s_1\ldots s_n ,s}$.
Given a $\Sigma$-equation $e$ of the form $l=r$.
Then $\Phi(e)$ is $\overline{\Phi}(l)=\overline{\Phi}(r)$ where
$\overline{\Phi}: T_{\Sigma}(X)\rightarrow T_{\Sigma'}(X')$ and $X'=\Phi(X)$.
An important property of these translations on algebras and equations
under signature morphisms is called \emph{satisfaction condition}, which
expresses the invariance of satisfaction under change of notation:
\begin{theorem}\textbf{\emph{(Satisfaction Condition~\cite{Gog94}).}}
\label{satisf}
Given a signature morphism $\Phi: \Sigma \rightarrow \Sigma'$,
a $\Sigma'$-algebra $A'$ and a $\Sigma$-equation $e$, %then:\\
$\Phi(A')\models^{\Sigma} e \mbox{ iff } A'\models^{\Sigma'} \Phi(e)$.
\end{theorem}

An  \textit{observational   signature}  is  a   many-sorted  signature
$\Sigma=(S,S_{obs},F)$  where  $S_{obs}\subseteq S$  is  the set  of
\textit{observable} sorts.  An \textit{Observational Specification} is
a  pair  $(\Sigma,E)$  where  $\Sigma$ is  an  observational
signature and $E$ is a set of axioms. We assume that axioms
are conditional equations with observable conditions.
A \emph{context} is a
term with exactly one occurrence of a distinguished variable, say $z$.
Observable contexts are contexts of observable sort. Let 
$C_{\Sigma}(s,s')$ be the set of contexts of sort $s'$ that contain a
distinguished variable of sort $s$. We write $c[t]$ for the replacement of
distinguished variable $z$ by the term $t$. A $\Sigma$-algebra $A$ 
\emph{behaviorally satisfies} an equation $l=r$, denoted $A\models^{\Sigma}_{obs} l=r$,
iff $A\models^{\Sigma} c[l]=c[r]$ for every observable context $c$. A model
of an observational specification $SP=(\Sigma,E)$ is a $\Sigma$-algebra $A$ that
behaviorally satisfies every axioms in $E$. We write $A\models^{\Sigma}_{obs} SP$ or
$A\models^{\Sigma}_{obs} E$. Also we write $E\models^{\Sigma}_{obs} e$ iff
$A\models^{\Sigma}_{obs} E$ implies $A\models^{\Sigma}_{obs} e$ where $e$ is
a (conditional)-equation.

\subsection{Component Specifications}\label{ccos}
Using Observational semantics we consider a Collaborative Object (CO) as a
\emph{black box} with a hidden (or non-observable) state~\cite{Gog00}. We only
specify the interactions between a user and an object. In the
following, we give our formalization:

\begin{definition}\textbf{\emph{(CO Signature).}}
Given        $S$       the        set       of        all       sorts,
$S_{b}=\{\emph{\texttt{State}},\emph{\texttt{Meth}}\}$ is  the set of
\emph{basic  sorts}  and  $S_{d}=S\setminus  S_{b}$ is  the  set  of
\emph{data sorts}.
  A \emph{CO  signature} $\Sigma = (S,S_{obs},F)$ is an  observational 
signature where the sort \emph{\texttt{State}} is the unique non-observable
sort.  The set of function symbols $F$ is defined as follows:
\begin{enumerate}[(1)]
%\begin{inparaenum}[(i)]
\item $F_{\emph{\texttt{Meth}}\,
      \emph{\texttt{State}},\emph{\texttt{State}}}=\{Do\}$, 
      $F_{\emph{\texttt{Meth}}\ \emph{\texttt{Meth}},
      \emph{\texttt{Meth}}}=\{IT\}$, $F_{\emph{\texttt{Meth}}\
      \emph{\texttt{State}},\emph{\texttt{Bool}}}=\{Poss\}$, and
      $F_{\omega,s}=\emptyset$ for all other cases where
      $\omega\in S^*_{b}$ and $s\in S_{b}$.

\item A function symbol $f : s_1\times s_2\times\ldots\times s_n \rightarrow 
      \emph{\texttt{Meth}}$ is called a \emph{method} if
      $s_1\cdot s_2\cdot \ldots \cdot s_n \in S^*_{d}$.
\item A function symbol $f : s_1\times s_2\times\ldots\times s_n \rightarrow s$
      is called an \emph{attribute} if:
\begin{inparaenum}[(i)]
\item $s_1\cdot s_2\cdot \ldots \cdot s_n$ contains only one 
      \emph{\texttt{State}} sort; and
\item $s\in S_{d}$.
\end{inparaenum}
\end{enumerate}
We use $\Sigma$, $\Sigma'$, $\Sigma_1$, $\Sigma_2$, $\ldots$, as variables
ranging over CO signatures. \hfill $\Box$
\end{definition}

The states of a collaborative object are accessible
using the function $Do$ which given a method and a state gives the resulting
state provided that the execution of this method is possible. For this we
use a boolean function $Poss$ that indicates the conditions under which a
method is enabled. The OT algorithm is denoted by the function symbol $IT$
which takes two methods as arguments and produces another method.

\begin{definition}\textbf{\emph{($\boldsymbol{\Sigma}$-Morphism).}}\label{def:mor}
  Given  CO  signatures  $\Sigma$   and  $\Sigma'$,  then  a  
$\Sigma$-\emph{morphism}  
$\Phi :  \Sigma  \rightarrow  \Sigma'$ is  a
  signature morphism such that:
%\begin{enumerate}
\begin{inparaenum}[(i)]
\item $\Phi(s)=s$ for all $s\in S_{d}$;
\item $\Phi(f)=f$ for all $f\in \Sigma_{\omega,s}$ where
      $\omega\in S^*_{d}$ and $s\in S_{d}$;
\item $\Phi(S_{b})=S'_{b}$ (where
      $S'_{b}=\{\emph{\texttt{State'}},\emph{\texttt{Meth'}}\}$,
      $\Phi(\emph{\texttt{State}})=\emph{\texttt{State'}}$ and
      $\Phi(\emph{\texttt{Meth}})=\emph{\texttt{Meth'}}$).\hfill $\Box$
% \item if $f'\in \Sigma'_{\omega',s'}$ and some sort in $\omega'$ is
%       \emph{\texttt{State'}} then $f'=\Phi(f)$ for some
%       $f\in \Phi$.
\end{inparaenum}
%\end{enumerate}
\end{definition}

The three conditions stipulate that $\Sigma$-morphisms
preserve \texttt{State} sort, observable sorts and functions. 
%The fourth
%condition expresses an important feature in object paradigm, namely
%the \emph{encapsulation}. Indeed, no new methods or attributes can be 
%defined on any signature source.

\begin{definition}\textbf{\emph{(Collaborative Component Specification).}}
  A \emph{collaborative component specification}  is a tuple 
$\mathcal{C}=(\Sigma,M,A,T,E)$ where:
%\begin{enumerate}
\begin{inparaenum}[(i)]
\item $\Sigma$ is a CO signature;
\item   $M$   is  a   set   of   method symbols,   \it{i.e.}
   $M=\{m\,   |\,
  m\in\Sigma_{\omega,\emph{\texttt{Meth}}}\mbox{   and   }   \omega\in
  S^*_{d}\}$;
\item   $A$  is  a   set  of   attribute symbols,  \it{i.e.}
  $A=\{a\,  |\,a\in\Sigma_{\omega,s}$   where   $\omega$   contains   exactly   one
  \emph{\texttt{State}} sort and $s\in S^*_{d}\}$;
\item $T$ is the set of axioms corresponding to the transformation function;
\item $E$ is the set of all axioms.
\end{inparaenum}
%\end{enumerate}
We let $\mathcal{C}$, $\mathcal{C}'$, $\mathcal{C}_1$, $\mathcal{C}_2$,
$\ldots$, denote collaborative component specifications.\hfill $\Box$
\end{definition}

In the following, we assume that all used (conditional) equations are
universally quantified.

\begin{example}\label{ex:1}
% Let  \verb'Color' be a sort containing a set of ordered color 
% constants such that \verb'BLACK', \verb'BLUE', etc.
% The following CO specifications \verb'CCHAR', \verb'CNAT' and \verb'CCOLOR' models
% a memory cell which keeps respectively
% a character value, a natural number value and a color value:
The following component specification \verb'CCHAR' models a memory cell 
(or a buffer) which stores a character value:
\begin{footnotesize}
\begin{verbatim}
spec CCHAR =
sort: Char Meth State
opns: Do : Meth State -> State
      putchar : Char -> Meth
      getchar : State -> Char
      IT : Meth Meth -> Meth
      maxchar : Char Char -> Char
axioms:
  (1) getchar(Do(putchar(c),st)) = c;
  (2) IT(putchar(c1),putchar(c2)) = putchar(maxchar(c1,c2));
\end{verbatim}
\end{footnotesize}

\verb'CCHAR' has one method \verb'putchar' and one attribute \verb'getchar'.
Axiom \verb'(2)' gives how to transform two concurrent \verb'putchar' in
order to achieve the data convergence. For that, we use function 
\verb+maxchar+ that computes the maximum of two character values. Note we
could have used another way to enforce convergence.

As the previous specification \verb'CNAT' and \verb'CCOLOR' model a memory 
cell which stores respectively a natural number value and a color value:
\begin{footnotesize}
\begin{verbatim}
spec CNAT =
sort: Nat Meth State
opns: Do : Meth State -> State
      putnat : Nat -> Meth
      getnat : State -> Nat
      IT : Meth Meth -> Meth
      minnat : Nat Nat -> Nat
axioms:
  (1) getnat(Do(putnat(n),st)) = n;
  (2) IT(putnat(n1),putnat(n2)) = putnat(minnat(n1,n2));

spec CCOLOR =
sort:
  Color Meth State
opns:
  Do : Meth State -> State
  putcolor : Color -> Meth
  getcolor : State -> Color
  IT : Meth Meth -> Meth
axioms:
  (1) getcolor(Do(putcolor(cl),st)) = cl;
  (2) IT(putcolor(cl1),putcolor(cl1)) = putcolor(mincolor(cl1,cl2));
\end{verbatim}
\end{footnotesize}

To get data convergence we have used in \verb'CNAT' (resp. \verb'CCOLOR')
another function \verb'minnat' (resp. \verb'mincolor') that computes 
the minimum value.
The sorts \verb+Char+, \verb+Nat+ and \verb'Color' are built-in.\hfill $\Box$
\end{example}

For a concise presentation and without loss of generality, we shall omit
the observable-sorted arguments from methods and attributes. We could
suppose we have one function for each of its possible arguments. For
instance, method \verb+putchar(+$c$\verb+)+ may be replaced by 
\verb+putchar+$_c$ for every $c\in$ \verb+CHAR+.

\begin{definition}\textbf{\emph{(($M$,$A$)-Complete).}}
Given a component specification $\mathcal{C}= (\Sigma,M,A,T,E)$.
The set $E$ is ($M$,$A$)-\emph{complete} iff all equations involving
$M$ have the form $C \implies a(Do(m,x))=t$,
%\[ C \implies a(Do(m,x))=t\]
where $x$ is a variable of sort \emph{\texttt{State}}, $a\in A$, $m\in M$,
$t\in T_{\Sigma\setminus M}(\{x\})$ and $C$ is a finite set of visible
pairs $t_1 = t'_1$, $t_2 = t'_2$, $\ldots$, $t_n = t'_n$ where
$t_1$, $t'_1\in T_{\Sigma}(X)_{s_1}$, $t_2$, $t'_2\in T_{\Sigma}(X)_{s_2}$, $\ldots$,
$t_n$, $t'_n\in T_{\Sigma}(X)_{s_n}$.\hfill $\Box$
\end{definition}

In Example~\ref{ex:1}, component specification \verb+CCHAR+ is
($M$,$A$)-complete as the only axiom involving methods 
(\textit{i.e.}, axiom \verb+(1)+) has the required form.
\verb+CNAT+ and \verb+CCOLOR+ are also ($M$,$A$)-complete.
In the remaining of this paper, we restrict our intention to component
specification which are ($M$,$A$)-complete.

As a component specification has an observational signature with one
non-observable sort, \texttt{State}, then the observable contexts have
the following form: $a(Do(m_n,\ldots,Do(m_1,s))$ where $m_1$,$\ldots$,
$m_n$ are methods and $a$ is an attribute.

\begin{definition}\textbf{\emph{(Specification morphisms).}}\label{def:specmo}
Given two collaborative component specifications 
$\mathcal{C}= (\Sigma,M,A,T,E)$ and
$\mathcal{C}'= (\Sigma',M',A',T',E')$, a \emph{specification morphism}
$\Phi : \mathcal{C} \rightarrow \mathcal{C}'$ is a signature morphism
$\Phi : \Sigma \rightarrow \Sigma'$ such that:
\begin{inparaenum}[(i)]
\item $\Phi(M)\subseteq M'$;
\item $\Phi(A)\subseteq A'$;
\item $E' \models^{\Sigma'}_{obs} \Phi(e)$ for each $e\in E$.\hfill $\Box$
\end{inparaenum}
\end{definition}

Definition~\ref{def:specmo} provides a support for reusing component
specification through the notion of specification morphism. Moreover,
it exploits the fact that the source component specification is
($M$,$A$)-complete by only requiring the satisfaction of finite number
of equations (see condition $(iii)$).
Note that Definitions~\ref{def:mor} and~\ref{def:specmo} have been used
for defining the \emph{static composition} that enables us to build up a composite
object from a \emph{fixed} number of other collaborative objects~\cite{Imi07}.
For instance, \verb+SIZEDCHAR+ is the composition of \verb+CCHAR+ and \verb+CNAT+
denoted by \verb+SIZEDCHAR+ = \verb+CCHAR+$\;\oplus\;$\verb+CNAT+.
This composition may be associated to an object with a character value and
an attribute for modifying the font size.
Due to limited space, the reader is referred to~\cite{Imi07}
for more details.

\subsection{Convergence Properties}\label{cp}
Before stating the properties that a component specification
$\mathcal{C}= (\Sigma,M,A,T,E)$ has to satisfy for ensuring convergence, we
introduce some notations.
Let $m_1$, $m_2$, \ldots, $m_n$ and $s$ be terms of sorts \texttt{Meth} and
\texttt{State} respectively:
\begin{enumerate}
\item applying a method sequence on a state is denoted as:
%\begin{gather*}
%(st)\lambda \overset{def}{=} st \\
\[
(s)[m_1;m_2;\ldots;m_n]~\triangleq~Do(m_n,\ldots,Do(m_2,Do(m_1,s))\ldots)
\]
%\end{gather*}
\item $Legal([m_1; m_2; \ldots; m_n],s)\triangleq
      Poss(m_1,s)\,\wedge\, 
      Poss(m_2, (s)m_1)\,\wedge\, \ldots$
      $\wedge\, Poss(m_n,(s)[m_1; m_2;\ldots; m_{n-1}])$.
\item $IT^*(m,[])=m$ and 
     $IT^*(m,[m_1; m_2;\ldots; m_{n-1}])=IT^*(IT(m,m_1),[m_2;\ldots; m_{n-1}])$
     where $[]$ is an empty method sequence.
\end{enumerate}

$\mathbf{TP1}$ expresses a state identity between two method sequences.
As mentioned before, we use an observational approach for
comparing two states. Accordingly, we define the condition $\mathbf{TP1}$
by the following state property (where the variables $st$, $m_1$ and
$m_2$ are universally quantified):
\begin{equation*}
\begin{array}{lcl}
CP1   & \triangleq & (Legal(seq_1,s)=true\; \wedge 
                         Legal(seq_2,s)=true) \implies (s) seq_1 = (s) seq_2
\end{array}
\end{equation*}
where $seq_1=[m_1; IT(m_2,m_1)]$ and $seq_2=[m_2; IT(m_1,m_2)]$.

Let $M' \subseteq M$ be a set of methods, we denote $CP1|_{M'}$ as
the restriction of $CP1$ to $M'$. Let $M_1, M_2\subseteq M $ be two
disjoint sets of methods, we define $CP1|_{M_1, M_2}$ as:
\begin{equation*}
\begin{array}{lcl}
CP1|_{M_1, M_2}   & \triangleq & (Legal(seq_i,s)=true\; \wedge
                                Legal(seq_j,s)=true) \implies (s) seq_i = (s) seq_j
\end{array}
\end{equation*}
where $seq_i=[m_i; IT(m_j,m_i)]$ and $seq_j=[m_j; IT(m_i,m_j)]$
such that $m_i \in M_i$ and $m_j \in M_j$ for all $i\neq j\in\{1,2\}$.

$\mathbf{TP2}$ stipulates a \textit{method identity} between two equivalent
sequences. Given three methods $m_1$, $m_2$ and $m_3$, transforming
$m_3$ with respect to two method sequences $[m_1; IT(m_2,m_1)]$ and
$[m_2; IT(m_1,m_2)]$ must give the same method.
We define $\mathbf{TP2}$ by the following property:
%(where $m_1$, $m_2$ and $m_3$
%are universally quantified):
\begin{gather*}
CP2 \, \triangleq IT^* (m_3, [m_1; IT(m_2,m_1)]) =
              IT^* (m_3, [m_2; IT(m_1,m_2)])
\end{gather*}

Let $M' \subseteq M$ be a set of methods, we denote $CP2|_{M'}$ as
the restriction of $CP2$ to $M'$. Let $M_1, M_2\subseteq M $ be two
disjoint sets of methods, we define $CP2|_{M_1, M_2}$ as:
\begin{gather*}
CP2|_{M_1, M_2} \, \triangleq
IT^* (m, [m'; IT(m'',m')]) =
IT^* (m, [m''; IT(m',m'')])
\end{gather*}

such that $m' \in M_i$, $m'' \in M_j$ and $m \in M_k$
for all $i,j,k\in\{1,2\}$  with $k\neq i$ or $k\neq j$.

The following definition gives the conditions under which a component
specification ensures the data convergence:

\begin{definition}\textbf{\emph{(Consistency).}}\label{def:cons}
$\mathcal{C}$ is said \emph{consistent} iff
$\mathcal{C} \models_{obs} CP1 \wedge CP2$.
\end{definition}

\section{Dynamic Composition}\label{sec:dc}
In this section, we present a construction that enables us to combine an arbitrary
number of the same collaborative object according to a given structure (we will
call it \emph{composition pattern}). 
In other words,
such objects are created and deleted dynamically. Thus, the obtained object has no
static structure.

\subsection{Basic Definitions}

\begin{definition}\textbf{\emph{(Composition Pattern).}}\label{comp_pat}
A \emph{composition pattern} is a parametric specification
$\overline{\mathcal{C}} = (PA,\mathcal{C})$ where :
\begin{itemize}
\item $PA = (\Sigma_{PA},E_{PA})$, called \emph{formal parameter}, is an algebraic
specification;
\item $\mathcal{C}=(\Sigma,M,A,T,E)$, called \emph{body}, is a collaborative 
component specification (or a component);
\end{itemize}
such that the following conditions hold:
%\begin{enumerate}
\begin{inparaenum}[(i)]
\item $S_{PA} =\{$\verb+Elem+$,$\verb+Bool+$\}$;
\item $\Sigma_{PA}\subset \Sigma$;
\item $E_{PA}\subset E$;
\item there exists a method symbol $m\in M$ containing at least one argument 
of \verb+Elem+ sort; this method is called \emph{parametric method}
\item there exists an attribute symbol $a\in A$ such that either its result is
of \verb+Elem+ sort or one of its  arguments is \verb+Elem+ sort; $a$ is called
\emph{parametric attribute}.
\end{inparaenum}
%\end{enumerate}

We let
$\overline{\mathcal{C}}$, $\overline{\mathcal{C}}'$, $\overline{\mathcal{C}}_1$,
$\overline{\mathcal{C}}_2$, $\ldots$, denote the composition patterns.
\hfill $\Box$
\end{definition}

%\noindent{\textbf{Remark.}} From the above definition we can deduce that
%$Elem\prec \texttt{State}_{\mathcal{C}}$.

\begin{example}\label{ex:pat}
The composition pattern \verb+PSET+ $= (PA,\mathcal{C})$ describes finite sets
with parametric element:

%dans le paramètre formel nous avons enlevé l'équation "eq(x,x)=true"
%pour éviter d'utiliser une égalité stricte et pouvoir utiliser
%une égalité observationnelle.
\noindent\emph{Formal parameter} $PA$ gives the properties of parameter sort
\verb+Elem+:

\begin{small}
\begin{boxedverbatim}
spec PA =
sorts: 
 Elem Bool
opns: 
 eq : Elem Elem -> Bool
axioms:
 (1) eq(x,y)=eq(y,x);
 (2) eq(x,y)=true, eq(y,z)=true => eq(x,z)=true;
\end{boxedverbatim}
\end{small}

\noindent\emph{Body} $\mathcal{C}$ is collaborative object reprsenting data set
of \verb+Elem+ sort:

\begin{small}
\begin{boxedverbatim}
spec C =
sorts: 
 Set Elem Bool
opns:
 empty : -> Set
 Do : Meth Set -> Set
 nop : -> Meth
 add : Elem -> Meth
 remove : Elem -> Meth
 Poss : Meth Set -> Bool
 iselem : Elem Set -> Bool
 IT : Meth Meth -> Meth
axioms:
 (1)  Poss(nop,st)=true;
 (2)  Poss(add(x),st)=true;
 (3)  iselem(x,st)=true  => Poss(remove(x),st)=true;
 (4)  iselem(x,st)=false => Poss(remove(x),st)=false;
 (5)  eq(x,y)=true  => iselem(x,Do(add(y),st))=true;
 (6)  eq(x,y)=false => iselem(x,Do(add(y),st))=iselem(x,st);
 (7)  eq(x,y)=true  => iselem(x,Do(remove(y),st))=false;
 (8)  eq(x,y)=false => iselem(x,Do(remove(y),st))=iselem(x,st);
 (9)  eq(x,y)=true  => IT(add(x),add(y))=nop;
 (10) eq(x,y)=false => IT(add(x),add(y))=add(x);
 (11) IT(add(x),remove(y))=add(x);
 (12) eq(x,y)=true  => IT(remove(x),remove(y))=nop;
 (13) eq(x,y)=false => IT(remove(x),remove(y))=remove(x);
 (14) IT(remove(x),add(y))=remove(x);
\end{boxedverbatim}
\end{small}
\hfill $\Box$
\end{example}

%\end{boxedverbatim}
%\end{small}

%\begin{small}
%\begin{boxedverbatim}

%%%
% \begin{example}\label{ex:pat}
% The composition pattern \verb+SET+ $= (PA,\mathcal{C})$ describes properties of finite 
% sets which are parameterized with respect to their elements:\\

% \noindent\emph{Formal parameter} $PA$:%\vspace{-2mm}
% \begin{verbatim}
% spec PA =
% sorts: 
%  Elem Bool
% cons: 
%  const : -> Elem
% opns:
%  eq : Elem Elem -> Bool
% axioms:
%  (1) eq(x,x)=true;
%  (2) eq(x,y)=eq(y,x);
%  (3) eq(x,y)=true, eq(y,z)=true => eq(x,z)=true;
% \end{verbatim}

% \noindent\emph{Body} $\mathcal{C}$:%\vspace{-2mm}
% \begin{verbatim}
% spec C =
% sorts: 
%  Set
% cons:
%  empty : -> Set
%  Do : Meth Set -> Set
%  nop : -> Meth
%  add : Elem -> Meth
%  remove : Elem -> Meth
% opns:
%  Poss : Meth Set -> Bool
%  iselem : Elem Set -> Bool
%  IT : Meth Meth -> Meth
% axioms:
%  (1)  Poss(nop,st)=true;
%  (2)  Poss(add(x),st)=true;
%  (3)  iselem(x,st)=true  => Poss(remove(x),st)=true;
%  (4)  iselem(x,st)=false => Poss(remove(x),st)=false;
%  (5)  iselem(x,empty)=false;
%  (6)  eq(x,y)=true  => iselem(x,Do(add(y),st))=true;
%  (7)  eq(x,y)=false => iselem(x,Do(add(y),st))=iselem(x,st);
%  (8)  eq(x,y)=true  => iselem(x,Do(remove(y),st))=false;
%  (9)  eq(x,y)=false => iselem(x,Do(remove(y),st))=iselem(x,st);
%  (10) eq(x,y)=true  => IT(add(x),add(y))=nop;
%  (11) eq(x,y)=false => IT(add(x),add(y))=add(x);
%  (12) IT(add(x),remove(y))=add(x);
%  (13) eq(x,y)=true  => IT(remove(x),remove(y))=nop;
%  (14) eq(x,y)=false => IT(remove(x),remove(y))=remove(x);
%  (15) IT(remove(x),add(y))=remove(x);
% \end{verbatim}
% \hfill $\Box$
% \end{example}
%%%

In the following definition, we give under which conditions a collaborative
component can substitute a formal parameter in a composition pattern.

\begin{definition}\textbf{\emph{(Admissibility).}}\label{def_adm}
Given $\overline{\mathcal{C}} = (PA,\mathcal{C})$ a composition
and $\mathcal{C}_1 = (\Sigma_1,M_1,A_1,T_1,E_1)$ a component
such that $(\Sigma \setminus \Sigma_{PA})\cap \Sigma_1 = \emptyset$
(\textit{i.e.} no similar names).
Component $\mathcal{C}_1$ is said \emph{admissible} for
$\overline{\mathcal{C}}$
if for all axioms $e\in E_{PA}$, $E_1  \models_{obs} \Phi(e)$, where
$\Phi :  \Sigma_{PA}  \rightarrow  \Sigma_1$ is a signature morphism
with $\Phi(\emph{\texttt{Elem}})=\emph{\texttt{State}}_{\mathcal{C}_1}$ and
$\Phi(\emph{\texttt{Bool}})=\emph{\texttt{Bool}}$. \hfill $\Box$
\end{definition}

Consider the character component \verb+CCHAR+ $=(\Sigma_1,M_1,A_1,T_1,E_1)$
given in Example~\ref{ex:1}. This component is admissible for the pattern
\verb+PSET+ (see Example~\ref{ex:pat}) by using the following morphism:
$\Phi(\texttt{Elem}) = \texttt{State}_{\texttt{CCHAR}}$ and
$\Phi(\texttt{eq}) = (=^{\texttt{CCHAR}}_{obs})$. This enables us to build up
a set of characters.

Substituting a formal parameter by an admissible component enables us to
build a new component.

\begin{definition}\textbf{\emph{(Instantiation parameter).}}\label{def_pp}
Let $\overline{\mathcal{C}_1} = (PA,\mathcal{C}_1)$ be a composition pattern.
Given $\mathcal{C}_2 = (\Sigma_2,M_2,A_2,T_2,E_2)$ an admissible component for
$\overline{\mathcal{C}_1}$ via a signature morphism
$\Phi :  \Sigma_{PA}  \rightarrow  \Sigma_2$.
The \emph{instantiation} of $\overline{\mathcal{C}_1}$ by $\mathcal{C}_2$, 
denoted by $\overline{\mathcal{C}_1}[PA \leftarrow \mathcal{C}_2]_{\Phi}$, is
the specification $(\Sigma,M,A,T,E)$ such that:
\begin{inparaenum}[(i)]
\item $\Sigma = \Sigma_2 \cup \Phi(\Sigma_{PA}) \cup (\Sigma \setminus \Sigma_{PA})$ ;
\item $M = \Phi(M_1)$ ;
\item $A = \Phi(A_1)$ ;
\item $T = \Phi(T_1)$ ;
\item $E = E_2 \cup \Phi(E_1)$.\hfill $\Box$
\end{inparaenum}
\end{definition}

% \noindent{[[[\textbf{Remark:}}: Don't forget to introduce the import relation.
% According to Definition~\ref{def_pp}, after instantiation we get
% $\prec = \prec_{\Phi(\mathcal{C}_1)} \cup \prec_{\mathcal{C}_2}$. ]]]

Although the below definition (see Definition~\ref{def_dc}) may seems 
rather complicated to understand,
it is just a mathematical formulation of some simple ideas how to build
a complex component -- with dynamic structure -- from a composition pattern
$\overline{\mathcal{C}_1}$ and an admissible component  $\mathcal{C}_2$:
\begin{itemize}
\item The formal parameter of $\overline{\mathcal{C}_1}$ is replaced by
  an admissible component $\mathcal{C}_2$ in order to build a new component
  $\mathcal{C}$.
\item This new component $\mathcal{C}$ is extended by a new method \texttt{Update}
whose role is to connect the $\overline{\mathcal{C}_1}$'s state space with
the $\mathcal{C}_2$'s state space. In other words, the use of \texttt{Update} means
that changing the state of $\mathcal{C}_2$ implies changing the state of 
$\overline{\mathcal{C}_1}$.  
\item Axioms given in $(iv)$ show how to transform \texttt{Update}. On the one hand,
we have to add axioms to define how to transform \texttt{Update} against other methods
of $\overline{\mathcal{C}_1}$. On the other hand, when modifying the same object
of $\mathcal{C}_2$ we use the transformation function related to $\mathcal{C}_2$.
But, the modification of two distinct objects of $\mathcal{C}_2$ can be performed
in any order (there is no interference).
\item Axioms given in $(v)$ state how attributes are altered by the method
\texttt{Update}.
\end{itemize}

\begin{definition}\textbf{\emph{(Dynamic Composition).}}\label{def_dc}
Given a composition pattern $\overline{\mathcal{C}_1} = (PA,\mathcal{C}_1)$,
a component $\mathcal{C}_2 = (\Sigma_2,M_2,A_2,T_2,E_2)$ and
a signature morphism $\Phi :  \Sigma_{PA}  \rightarrow  \Sigma_2$.
Let
$Update : s_1\ldots s_n\;\emph{\texttt{State}}_{\mathcal{C}_2}\;
\emph{\texttt{State}}_{\mathcal{C}_2}\;
\rightarrow \emph{\texttt{Meth}}$ be a method symbol.
The specification
$\mathcal{C} = (\Sigma,M,A,T,E)$ is said 
a \emph{dynamic composition} of $\mathcal{C}_2$ with respect to
$\overline{\mathcal{C}_1}$ (denoted $\overline{\mathcal{C}_1}[\mathcal{C}_2]$) 
iff $\mathcal{C}_2$ is admissible for 
$\overline{\mathcal{C}_1}$ via $\Phi$, and 
$\mathcal{C} = \overline{\mathcal{C}_1} [PA \leftarrow
\mathcal{C}_2]_{\Phi} \cup (\Sigma',M',A',T',E')$ such that:
\begin{enumerate}[(i).]
\item $\Sigma'=(S',F')$ with $S' = S_2 \cup \Phi(S_1)$ and
      $F' = \{\emph{\texttt{Update}}\}$. Method $\emph{\texttt{Update}}(U,x,y)$ means the replacement 
of the old value
       $x$ by the new one $y$. The value $y$ is considered as the result given by
applying a method of $\mathcal{C}_2$ on $x$ ($U$ denotes a sequence of variables
$x_1$, $\ldots$, $x_n$).
\item $M' = \{\emph{\texttt{Update}}(U,x,y)\, |\, x,y$ are variables of sort
      $\emph{\texttt{State}}_{\mathcal{C}_2}$ and $U$ is a variable of sort
      $S_d^*\}$;
\item $A' = \emptyset$;
\item Le $u_1 = \emph{\texttt{Update}}(U,x,Do_{\mathcal{C}_2}(m_1,x))$ and
      $u_2 = \emph{\texttt{Update}}(U',x',Do_{\mathcal{C}_2}(m_2,x'))$ be two methods where
      $m_1$, $m_2\in \mathcal{C}_2$. 
      For every method $m \in \Phi(M_1)$, we have:
\[
      T' = \textbf{Ax}(IT(u_1,m))\; \cup
      \textbf{Ax}(IT(m,u_1))\; \cup
      \textbf{Ax}(IT(u_1,u_2))) 
\]
such that $\textbf{Ax}(IT(u_1,u_2))$ contains the following axioms:
\begin{gather*}
U = U' \wedge x=x' \implies IT(u_1,u_2)=u'_1\\
x \neq x ' \implies IT(u_1,u_2) = u_1\\
U \neq U' \implies IT(u_1,u_2) = u_1
\end{gather*}
with $u'_1 = \emph{\texttt{Update}}(U, Do_{\mathcal{C}_2}(m_2,x'),
Do_{\mathcal{C}_2}(IT_{\mathcal{C}_2}(m_1,m_2),$
$Do_{\mathcal{C}_2}(m_2,x')))$.
\item For each attribute symbol $a: s'_1\ldots s'_m \rightarrow s'$,
      %$a \in \Phi(A_1)$, 
      we have
\[    E' =  \textbf{Ax}(Poss( \emph{\texttt{Update}}(U,x,y),st)) \cup 
      \textbf{Ax}(a(Z,Do( \emph{\texttt{Update}}(U,x,y),st))) \]
     where $\textbf{Ax}(a(Z,Do( \emph{\texttt{Update}}(V,x,y),st)))$ is defined as follows:
     \begin{enumerate}
     \item $a$ is the instance of a parametric attribute whose one of its arguments
is of sort $\Phi(Elem)$:
\begin{gather*}
C[Z,x',U,x,y,st] \implies a(Z,x',Do( \emph{\texttt{Update}}(U,x,y),s)) = cst\\
\overline{C[Z,x',U,x,y,st]} \implies a(Z,x',Do( \emph{\texttt{Update}}(U,x,y),st)) = a(Z,x',st)
\end{gather*}
           with $cst$ is constant of sort $s'$ and $C[Z,x',V,x,y,st]$   
           ($\overline{C[Z,x',V,x,y,st]}$ is its negation) is a formula (containing
free variables) built up of conjunction of observable equations in such a way that
           $C[Z,x',U,x,y,st] \wedge C[Z,x',U',x,y,st]$ is false whenever
           $U\neq U'$.
     \item $a$ is the instance of a parametric attribute with $s' = \Phi(Elem)$:
\begin{gather*}
C'[Z,U,st] \implies a(Z,Do( \emph{\texttt{Update}}(U,x,y),st)) = y\\
\overline{C'[Z,U,st]} \implies a(Z,Do( \emph{\texttt{Update}}(U,x,y),st)) = a(Z,st)
\end{gather*}
          where $C'[Z,U,st]$ (and its negation) is a formula (containing
free variables)  built up of conjunction of observable equations in such a way that
          $C'[Z,U,st] \wedge C'[Z,U',st]$ is false whenever $U\neq U'$.
     \item $a$ is not the instance of a parametric attribute:
          $a(Z,Do(\emph{\texttt{Update}}(U,x,y),st)) = a(U,st)$.
     \end{enumerate}
\end{enumerate}
The notation $\textbf{Ax}(f)$ means the set of axioms used for defining function
$f$. \hfill $\Box$
\end{definition}

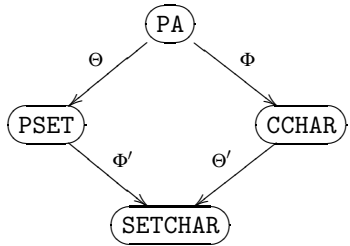
\begin{figure}[htb]
\begin{small}
\centerline{
\xymatrix@C=10pt@M=2pt@R=10pt{
& &
*++[F-:<10pt>]\txt{\texttt{PA}} \ar[ddl]_{\Theta} \ar[ddr]^{\Phi} & &\\
& \\
& *++[F-:<10pt>]\txt{\texttt{PSET}} \ar[ddr]^{\Phi'} & & 
*++[F-:<10pt>]\txt{\texttt{CCHAR}} \ar[ddl]_{\Theta'}\\
& \\
& &*++[F-:<10pt>]\txt{\texttt{SETCHAR}} & &\\
}}
\end{small}
\label{fig:dyncomp}
\caption{Dynamic Composition.}
\end{figure}

\begin{example} Figure~\ref{fig:dyncomp} shows
the dynamic composition of \verb+CCHAR+ (see Example~\ref{ex:1})
with respect to
\verb+PSET+ (see Example~\ref{ex:pat}), via the following morphism
$\Phi(Elem) = \texttt{State}_{\texttt{CCHAR}}$ and
$\Phi(\texttt{eq}) = (=^{\texttt{CCHAR}}_{obs})$. 
Note that $\Theta$ and $\Theta'$ are only inclusion morphisms~\cite{Wir90,Gog00}.
The composition proceeds by the following steps:
\begin{enumerate}
\item The instantiation of  \verb+PSET+ via $\Phi$, \textit{i.e.}   
      \verb+SETCHAR+ = $\Phi(\texttt{PSET})$ ;
\item Add to \verb+SETCHAR+ a new method 
      $\emph{\texttt{Update}} : \texttt{State}_{\texttt{CHAR}}\; 
      \texttt{State}_{\texttt{CHAR}} \rightarrow \texttt{Meth}$ with
      the following axioms:
      \begin{enumerate}
      \item Transforming \emph{\texttt{Update}} methods
      (see Definition~\ref{def_dc}.($iv$)):
\begin{verbatim}
(16) c1 = c2  => IT(Update(c1,c2),Update(c3,c4)) = Update(c4,c')
(17) c1 <> c2 => IT(Update(c1,c2),Update(c3,c4)) = Update(c1,c2)
\end{verbatim}
           where \verb+c'=Do_CCHAR(IT_CCHAR(m1,m2),c4)+, \verb+m1+ and
           \verb+m2+ are methods of  \verb+CCHAR+ such that
           \verb+c2 = Do_CCHAR(m1,c1)+ and \verb+c4 = Do_CCHAR(m2,c3)+.
      \item Axioms for defining function $Poss$:
\begin{verbatim}
(18) iselem(c,st)=true  => Poss(Update(c,c'),st) = true
(19) iselem(c,st)=false => Poss(Update(c,c'),st) = false
\end{verbatim}
      \item axioms for all attributes observing the effects of \emph{\texttt{Update}}
            (see Definition~\ref{def_dc}.($v$)) :
\begin{verbatim}
(20) c = c2  => iselem(c,Do(Update(c1,c2),st)) = true
(21) c1 <> c2 => iselem(c,Do(Update(c1,c2),st)) = iselem(c,st)
\end{verbatim}
      \end{enumerate}
\end{enumerate}\hfill $\Box$
\end{example}

\subsection{Illustrative Example}
In  word processor softwares (such as MicroSoft Word),
a document has a hierarchical structure.
It contains not only text but also formatting objects
(font, color, size, etc). Typically,
a document is divided into pages, paragraphs,
phrases, words and characters. A formatting object may be found
in each of these levels. Several collaborative editors rely on this
document structure, as  CoWord~\cite{Sun06} that is a collaborative version
of MicroSoft Word.
Now we will present how to model this document structure
using a dynamic composition. Note that each level has a linear structure,
except of characters. So, we use a composition pattern \verb+STRING+ that
represents a sequence of elements. The formal parameter \verb+Elem+ of
\verb+STRING+ can be substituted by any component.
Moreover, this pattern has two methods:
\begin{inparaenum}[(i)]
\item \verb+Ins(p,e,n)+ to add element \verb+e+ at position \verb+p+;
\item \verb+Del(p,n)+ to remove the element at at position \verb+p+.  
\end{inparaenum}
The argument \verb+n+ is the identity of the issuer (user or) site.

Suppose we want to equip the
document with formatting objects such as the size and color. So, consider
the components \verb+CCHAR+ (a character component), \verb+CNAT+ (a size
component) and \verb+CCOLOR+ (a color component) (see Example~\ref{ex:1}).
The basic element in our structure document is the formatted character (an
object character with color and size attributes),
\verb+FCHAR+ that is obtained by a static composition~\cite{Imi07}:
\verb+FCHAR+ = \verb+CCHAR+$\;\oplus\;$\verb+CNAT+$\;\oplus\;$\verb+CCOLOR+.

A formatted word is a sequence of formatted characters that is built 
up by dynamic and static compositions:
\verb+WORD+ = \verb+STRING+$[$\verb+FCHAR+$]$ and 
\verb+FWORD+ = \verb+WORD+$\;\oplus\;$\verb+CNAT+$\;\oplus\;$\verb+CCOLOR+.

%% \begin{center}
%% \verb+WORD+ = \verb+STRING+$[$\verb+FCHAR+$]$ and 
%% \verb+FWORD+ = \verb+WORD+$\;\oplus\;$\verb+CNAT+$\;\oplus\;$\verb+CCOLOR+
%% \end{center}

The remaining levels are built up in the same way:
\begin{center}
\verb+SENTENCE+ = \verb+STRING+$[$\verb+FWORD+$]$ and
\verb+FSENTENCE+ = \verb+SENTENCE+$\;\oplus\;$\verb+CNAT+$\;\oplus\;$\verb+CCOLOR+\\
\verb+PARAGRAPH+ = \verb+STRING+$[$\verb+FSENTENCE+$]$ and
\verb+FPARAGRAPH+ = \verb+PARAGRAPH+$\;\oplus\;$\verb+CNAT+$\;\oplus\;$\verb+CCOLOR+\\
\verb+PAGE+ = \verb+STRING+$[$\verb+FPARAGRAPH+$]$ and 
\verb+FPAGE+ = \verb+PAGE+$\;\oplus\;$\verb+CNAT+$\;\oplus\;$\verb+CCOLOR+
\end{center}

%Il est clair que la consistance est régie par le type de composition
%ainsi que les composants utilisés à chaque niveau.

\section{Correctness}\label{sec:cor}
In this section, we present the correctness of our dynamic composition by
enumerating the following properties.

\medskip
Applying \texttt{Update} on two distinct objects can be performed in any
order.

\begin{lemma}\label{dcom1}
Let $a : s_1\ldots s_n\; \emph{\texttt{State}}\rightarrow s$ be an attribute
such that $a$ is the instance of a parametric attribute. Given two methods
$u_1 = \emph{\texttt{Update}}(U,x,x')$ and
$u_2 = \emph{\texttt{Update}}(V,y,y')$. If $U\neq V$ or $x\neq y$ then:\\
$a(Z,(st)[u_1 ; u_2]) = a(Z,(st)[u_2 ; u_1])$   for all states $st$.
\hfill $\Box$
\end{lemma}

\noindent{\textbf{Proof.}}
Two cases are considered:\\
\noindent{\textit{First case:}} there is only one argument
$s_i =\Phi(Elem) = \texttt{State}_{\mathcal{C}_2}$ with
$i\in\{1,\ldots,n\}$ such that:
$a : s_1\ldots s_{n-1} \texttt{State}_{\mathcal{C}_2}\;
\texttt{State}\rightarrow s$.
According to Definition~\ref{def_dc} we have:\\
%\begin{gather*}
$a(Z,z,(st)[\texttt{Update}(U,x,x') ; \texttt{Update}(V,y,y')]) = 
a(Z,z,(st)[\texttt{Update}(V,y,y') ; \texttt{Update}(U,x,x')])$
%\end{gather*}
\begin{enumerate}
\item $U = V$ and $x\neq y$:
      \begin{enumerate}
      \item if $C[Z,z,U,x,x',st]\wedge C[Z,z,V,y,y',st]$ is true then $cst = cst$;
      \item if $C[Z,z,U,x,x',st]\wedge \overline{C[Z,z,V,y,y',st]}$
            is true then $cst = cst$;
      \item if $\overline{C[Z,z,U,x,x',st]}\wedge C[Z,z,V,y,y',st]$
            is true then $cst = cst$;
      \item if $\overline{C[Z,z,U,x,x',st]}\wedge \overline{C[Z,z,V,y,y',st]}$ 
            is true than $a(Z,z,st) = a(Z,z,st)$;
      \end{enumerate}

\item $U\neq V$: According to Definition~\ref{def_dc} we have
      $C[Z,z,U,x,y,st] \wedge C[Z,z,U',x,y,st]$ is false whenever that
      $U\neq U'$. Three cases are possible:
      \begin{enumerate}
      \item if $C[Z,z,U,x,x',st]\wedge \overline{C[Z,z,V,y,y',st]}$
            is true then $cst = cst$;
      \item if $\overline{C[Z,z,U,x,x',st]}\wedge C[Z,z,V,y,y',st]$
            is true then $cst = cst$;
      \item if $\overline{C[Z,z,U,x,x',st]}\wedge \overline{C[Z,z,V,y,y',st]}$ 
            is true then $a(Z,z,st) = a(Z,z,st)$;
      \end{enumerate}
\end{enumerate}

\noindent{\textit{Second case:}} 
$s = \Phi(Elem) = \texttt{State}_{\mathcal{C}_2}$ such that:
$a : s_1\ldots s_{n-1} \;
\texttt{State}\rightarrow \emph{\texttt{State}}_{\mathcal{C}_2}$.
According to Definition~\ref{def_dc} we get:
%\begin{gather*}
$a(Z,(st)[\texttt{Update}(U,x,x') ; \texttt{Update}(V,y,y')]) =
a(Z,(st)[\texttt{Update}(V,y,y') ; \texttt{Update}(U,x,x')])$
%\end{gather*}

\begin{enumerate}
\item $U = V$ and $x\neq y$ : as $u_1$ and $u_2$ are applied on state $st$ then
      $a(Z,st)=x$ and $a(Z,st)=y$. Thus, we have $x=y$ that is a contradiction
      of this case.
\item $U\neq V$: According to Definition~\ref{def_dc} we have
      $C'[Z,U,st] \wedge C'[Z,U',st]$ is false whenever
      $U\neq U'$. So, we have the following cases:
      \begin{enumerate}
      \item if $C'[Z,U,st]\wedge \overline{C'[Z,V,st]}$ is true then
            $x' = x'$;
      \item if $\overline{C'[Z,U,st]}\wedge C'[Z,V,st]$ is true then
            $x' = x'$;
      \item if $\overline{C'[Z,U,st]}\wedge \overline{C'[Z,V,st]}$ 
            is true then  $a(Z,st) = a(Z,st)$;\hfill $\Box$
      \end{enumerate}
\end{enumerate}

\medskip
If two \texttt{Update} methods $u_1$ and $u_2$ modify two distinct objects respectively
then both sequences $[u_1 ; u_2]$ and $[u_2 ; u_1]$ have the same effect.

\begin{lemma}\label{dcom2}
Let $u_1 = \emph{\texttt{Update}}(U,x,x')$ and $u_2 = \emph{\texttt{Update}}(V,y,y')$ be two methods.
For all states $st$, if  $U\neq V$ or $x\neq y$ then 
$(st)[u_1 ; u_2] =_{obs} (st)[u_2 ; u_1]$.
\hfill $\Box$
\end{lemma}

\noindent{\textbf{Proof.}}
Consider an arbitrary context $C[st] = a\cdot m_1\cdot\ldots\cdot m_n$
for $n>0$
with $a\in A$  and  $m_i\in M$ such that
$i\in\{1,\ldots,n\}$. Next we have: $C[(st)[u_1 ; u_2]] = C[(st)[u_2 ; u_1]]$.

It is sufficient to prove by induction on $n$ that:\\
$a(Z,(st)[u_1;u_2;m_1(X_1);\ldots; m_n(X_n)]) = 
a(Z,(st)[u_2;u_1;m_1(X_1);\ldots; m_n(X_n)])$.

\medskip
\noindent{\textit{Basis induction:}} For $n=0$ and $C[st] = a$ we have:
\begin{equation}\label{for:f1}
a(Z,(st)[u_1;u_2]) = a(Z,(st)[u_2;u_1]). 
\end{equation}
To prove Equation~(\ref{for:f1}) we have to consider two cases:
\begin{enumerate}[(i)]
\item $a$ is the instance of a parametric attribute:
      Equation~(\ref{for:f1}) is then true by using Lemma~\ref{dcom1}.
\item $a$ is not the instance of a parametric attribute: 
According to Definition~\ref{def_dc} we have
      $a(Z,(st)[u_1;u_2]) = a(Z,st)$ and $a(Z,(st)[u_2;u_1]) = a(Z,st)$.
\end{enumerate}

\medskip
\noindent{\textit{Induction hypothesis:}} For $n>0$
$a(Z,(st)[u_1;u_2;m_1(X_1);\ldots; m_n(X_n)]) = 
a(Z,(st)[u_2;u_1;m_1(X_1);\ldots; m_n(X_n)])$

\medskip
\noindent{\textit{Induction step:}}
We show now if $C'[st]=a\cdot m_1\cdot\ldots\cdot m_n\cdot m_{n+1}$ 
then $C[(st)[u_1 ; u_2]] = C[(st)[u_2 ; u_1]]$. 
Let $st_1 = (st)[u_1;u_2;m_1(X_1);\ldots; m_n(X_n)]$ and
$st_2 = (st)[u_2;u_1;m_1(X_1);\ldots; m_n(X_n)]$. By induction hypothesis
we deduce that $st_1 =_{obs} st_2$. 
As $=_{obs}$ is a congruence then
$a(Z,(st_1)[m_{n+1}]) = a(Z,(st_2)[m_{n+1}])$.\hfill $\Box$

\medskip
The dynamic composition of a consistent component with respect to a consistent
composition pattern produces a new component that satisfies $CP1$ for all
\texttt{Update} methods.

\begin{theorem}\label{dcom3}
Given a composition pattern $\overline{\mathcal{C}_1} = (PA,\mathcal{C}_1)$ 
and a component $\mathcal{C}_2 = (\Sigma_2,M_2,A_2,T_2,E_2)$. Let
$\mathcal{C} = (\Sigma,M,A,T,E)$ be the dynamic composition of
$\mathcal{C}_2$ with respect to $\overline{\mathcal{C}_1}$. If
$\overline{\mathcal{C}_1}$ and $\mathcal{C}_2$ are consistent then
$E\models_{obs} CP1\mid_{M'}$ with $M'$ is the set of \emph{\texttt{Update}} methods.
\hfill $\Box$
\end{theorem}

\noindent{\textbf{Proof.}}
$CP1\mid_{M'}$ is defined as follows:
\begin{gather*}
(st)[\texttt{Update}(X,u,v);IT(\texttt{Update}(Y,u',v'),\texttt{Update}(X,u,v))] = \\
(st)[\texttt{Update}(Y,u',v');IT(\texttt{Update}(X,u,v),\texttt{Update}(Y,u',v'))]
\end{gather*}
where $v = Do_{\mathcal{C}_2}(m_1(V),u)$ and $v' = Do_{\mathcal{C}_2}(m_2(W),u')$
with $m_1$ and $m_2$ are methods in $\mathcal{C}_2$. According to 
Definition~\ref{def_dc} we consider two cases:

\noindent{\textit{First case:}} $X=Y$ and $u=u'$

$CP1\mid_{M'}$ is rewritten as follows:
\begin{gather*}
(st)[\texttt{Update}(X,u,v);
\texttt{Update}(Y,v,Do_{\mathcal{C}_2}(IT_{\mathcal{C}_2}(m_2(W),m_1(V)),v))] = \\
(st)[\texttt{Update}(Y,u,v');
\texttt{Update}(X,v',Do_{\mathcal{C}_2}(IT_{\mathcal{C}_2}(m_1(V),m_2(W)),v'))]
\end{gather*}

As $v = Do_{\mathcal{C}_2}(m_1(V),u)$, $v' = Do_{\mathcal{C}_2}(m_2(W),u)$ and
$\mathcal{C}_2$ is consistent then
\[
Do_{\mathcal{C}_2}(IT_{\mathcal{C}_2}(m_2(W),m_1(V)),v) =
Do_{\mathcal{C}_2}(IT_{\mathcal{C}_2}(m_1(V),m_2(W)),v') = u''
\]

Thus we get:
$(st)[\texttt{Update}(X,u,v); \texttt{Update}(Y,v,u'')] = 
(st)[Update(Y,u,v');Update(X,v',u'')]$
that is true.

\noindent{\textit{Second Case:}} $X \neq Y$ or $u \neq u'$

$CP1\mid_{M'}$ is rewritten as follows:
\begin{gather*}
(st)[\texttt{Update}(X,u,v);\texttt{Update}(Y,u',v')] = 
(st)[\texttt{Update}(Y,u',v');\texttt{Update}(X,u,v)]
\end{gather*}
This equation is always true according to Lemma~\ref{dcom2}.
\hfill $\Box$

\medskip
The dynamic composition of a consistent component with respect to a consistent
composition pattern produces a new component that satisfies $CP2$ for all
\texttt{Update} methods.

\begin{theorem}\label{dcom4}
Given a composition pattern $\overline{\mathcal{C}_1} = (PA,\mathcal{C}_1)$ 
and a component $\mathcal{C}_2 = (\Sigma_2,M_2,A_2,T_2,E_2)$. Let
$\mathcal{C} = (\Sigma,M,A,T,E)$ be the dynamic composition of
$\mathcal{C}_2$ with respect to $\overline{\mathcal{C}_1}$. If
$\overline{\mathcal{C}_1}$ and $\mathcal{C}_2$ are consistent then
$E\models_{obs} CP2\mid_{M'}$ with $M'$ is the set of \emph{\texttt{Update}} methods.
\hfill $\Box$
\end{theorem}

\noindent{\textbf{Proof.}}
Let $up \triangleq \texttt{Update}(R,v,w)$, $up_1 \triangleq \texttt{Update}(P,x,y)$ 
and $up_2 \triangleq \texttt{Update}(Q,z,t)$ be three methods, where
$w = Do_{\mathcal{C}_2}(m(Z),v)$, 
$y = Do_{\mathcal{C}_2}(m_1(V),x)$ and $t = Do_{\mathcal{C}_2}(m_2(W),z)$ with
$m$, $m_1$ and $m_2$ are methods in $\mathcal{C}_2$. 
Condition $CP2\mid_{M'}$ is defined as follows:
\begin{gather*}
IT^* (up, [up_1 ; IT(up_2,up_1)]) = IT^* (up, [up_2 ; IT(up_1,up_2)])
\end{gather*}
According to Definition~\ref{def_dc} we consider two cases:

\noindent{\textit{First case:}} $P=Q$ and $x=z$

$CP2\mid_{M'}$ is rewritten as  $IT^* (up, [up_1 ; up_2']) = IT^* (up, [up_2 ; up_1'])$
where:
\begin{gather*}
up_1' \triangleq  \texttt{Update}(P,Do_{\mathcal{C}_2}(m_2(W),z),
Do_{\mathcal{C}_2}(IT_{\mathcal{C}_2}(m_1(V),m_2(W)),Do_{\mathcal{C}_2}(m_2(W),z)))\\
\mbox{and}\\
up_2' \triangleq  \texttt{Update}(Q,Do_{\mathcal{C}_2}(m_1(V),x),
Do_{\mathcal{C}_2}(IT_{\mathcal{C}_2}(m_2(W),m_1(V)),Do_{\mathcal{C}_2}(m_1(V),x)))
\end{gather*}

Two cases are possible:
\begin{enumerate}
\item $R = P$ and $v=x$.
In this case we get:
\begin{gather*}
\texttt{Update}(R,u_1,Do_{\mathcal{C}_2}(m',u_1)) = 
\texttt{Update}(R,u_2,Do_{\mathcal{C}_2}(m'',u_2))\mbox{ where}\\
u_1 \triangleq Do_{\mathcal{C}_2}(IT_{\mathcal{C}_2}(m_2(W),m_1(V)),Do(m_1(V),x))\\
m' \triangleq IT^*_{\mathcal{C}_2}(m(Z),[m_1(V);IT_{\mathcal{C}_2}(m_2(W),m_1(V))])\\
u_2 \triangleq Do_{\mathcal{C}_2}(IT_{\mathcal{C}_2}(m_1(V),m_2(W)),Do(m_2(W),z))\\
m'' \triangleq IT^*_{\mathcal{C}_2}(m(Z),[m_2(W);IT_{\mathcal{C}_2}(m_1(V),m_2(W))])
\end{gather*}
Since $\mathcal{C}_2$ is consistent, then
$u_1 = u_2$ and $m' = m''$. Consequently, the above equation is true.
\item $R\neq P$ or $v\neq x$.
We have $IT^* (up, [up_1 ; up_2'])= up$ and $IT^* (up, [up_2 ; up_1'])=up$.
\end{enumerate}

\noindent{\textit{Second case:}} $P \neq Q$ or $x \neq z$
$CP2\mid_{M'}$ is rewritten as follows:
\begin{gather*}
IT^*(\texttt{Update}(R,v,w),[\texttt{Update}(P,x,y);\texttt{Update}(Q,z,t)]) =\\
IT^*(\texttt{Update}(R,v,w),[\texttt{Update}(Q,z,t);\texttt{Update}(P,x,y)])
\end{gather*}

Three cases are considered:
\begin{enumerate}
\item $R = P$ and $v=x$.
We get:
\begin{gather*}
\texttt{Update}(R,Do_{\mathcal{C}_2}(m_1(V),x),
Do_{\mathcal{C}_2}(IT_{\mathcal{C}_2}(m(Z),m_1(V)),Do(m_1(V),x))) =\\
\texttt{Update}(R,Do_{\mathcal{C}_2}(m_1(V),x),
Do_{\mathcal{C}_2}(IT_{\mathcal{C}_2}(m(Z),m_1(V)),Do(m_1(V),x)))
\end{gather*}

\item $R = Q$ and $v=z$.
We get:
\begin{gather*}
\texttt{Update}(R,Do_{\mathcal{C}_2}(m_2(W),z),
Do_{\mathcal{C}_2}(IT_{\mathcal{C}_2}(m(Z),m_2(W)),Do(m_2(W),z))) =\\
\texttt{Update}(R,Do_{\mathcal{C}_2}(m_2(W),z),
Do_{\mathcal{C}_2}(IT_{\mathcal{C}_2}(m(Z),m_2(W)),Do(m_2(W),z)))
\end{gather*}

\item $R\neq P$, $R\neq Q$, $v\neq x$ or $v\neq z$.
We get $\texttt{Update}(R,v,w) = \texttt{Update}(R,v,w)$.
\end{enumerate}
\hfill $\Box$

\medskip
The following theorem is very important since it stipulates that
the consistency property is preserved by dynamic composition.

\begin{theorem}\label{dcom5}
Given a consistent composition pattern $\overline{\mathcal{C}_1} = (PA,\mathcal{C}_1)$ 
and a consistent component $\mathcal{C}_2 = (\Sigma_2,M_2,A_2,T_2,E_2)$. Let
$\mathcal{C} = (\Sigma,M,A,T,E)$ be the dynamic composition
$\mathcal{C}_2$ with respect to $\overline{\mathcal{C}_1}$ via
the morphism $\Phi$. If
$E\models_{obs} CP1\mid_{M',\Phi(M_1)}$ and $E\models_{obs} CP2\mid_{M',\Phi(M_1)}$
then $\mathcal{C}$ is consistent where $M'$ is the set \emph{\texttt{Update}} methods.
\hfill $\Box$
\end{theorem}

\noindent{\textbf{Proof.}}
Assume that $E\models_{obs} CP1\mid_{M',\Phi(M_1)}$ and
$E\models_{obs} CP2\mid_{M',\Phi(M_1)}$. By definition, $\mathcal{C}$
is consistent iff
$E\models_{obs} CP1 \wedge CP2$.
\begin{enumerate}
\item Proof of $E\models_{obs} CP1$.
Condition $CP1$ can be expressed as follows:
\[
CP1 \triangleq CP1\mid_{M'} \wedge \Phi(CP1\mid_{M_1})
\wedge CP1\mid_{M',\Phi_2(M_1)}
\]
\vspace{-4pt}
As $\overline{\mathcal{C}_1}$ is consistent and according to Theorem~\ref{dcom3}
$CP1$ is then satisfied.
\item Proof of $E\models_{obs} CP2$. Condition $CP2$ can be given as follows:
\[
CP2 \triangleq CP2\mid_{M'} \wedge \Phi(CP2\mid_{M_1})
\wedge CP2\mid_{M',\Phi_2(M_1)}
\]
\vspace{-4pt}
Since $\overline{\mathcal{C}_1}$ is consistent then
$CP2$ is true (By Theorem~\ref{dcom4}). \hfill $\Box$
\end{enumerate}
%\hfill $\Box$

%% Il faut souligner que la composition dynamique $\mathcal{C}$ est
%% semi-consistante ssi :

%% \begin{enumerate}
%% \item $E\models_{obs} CP1\mid_{M',\Phi(M_1)}$ ;
%% \item $E\models_{obs} CP2\mid_{M',\Phi(M_1)}$ ;
%% \item le patron $\overline{\mathcal{C}_1}$ est semi-consistant ou/et
%%        le composant $\mathcal{C}_2$ est semi-consistant.
%% \end{enumerate}

\section{Conclusion}
In this work, we have proposed a formal component-based design
for composing collaborative objects. 
We have dealt with the composition of arbitrary number of collaborative objects by
using a dynamic composition in such a way the objects are created and
deleted dynamically.
Moreover, we have provided sufficient conditions for preserving $TP1$ and $TP2$
by the dynamic composition.

As future work, we intend to study the semantic properties of static and dynamic
compositions. 
Finally, we want to implement these compositions on top of the verification
techniques given in~\cite{Imi06,Han09}.%\vspace{-5mm}
%correct OT algorithms.

\bibliographystyle{eptcs} % or whatever you prefer
\bibliography{mybib,induc}

\end{document}